\documentclass[11pt,aps,pre,onecolumn,superscriptaddress,noshowpacs,longbibliography,floatfix]{revtex4-2}
\usepackage[utf8]{inputenc}
\usepackage{cancel}
\usepackage[left=0.9in,right=0.9in,top=0.9in,bottom=0.9in]{geometry}
\usepackage{setspace}
\usepackage{graphics}
\usepackage{amssymb}
\usepackage{xfrac}
\usepackage{amsmath}
\usepackage{bm}
\usepackage{mathtools}
\usepackage{times}
\usepackage{color}
\usepackage{xcolor}
\usepackage{graphicx}

\newcommand{\dee}{\,{\rm d}}
\newcommand{\calS}{{\mathcal S}}
\newcommand{\calO}{{\mathcal O}}

\newcommand{\calD}{{\mathcal D}}
\newcommand{\calJ}{{\mathcal J}}

\newcommand{\calP}{{\mathcal P}}
\newcommand{\eqa}{\begin{eqnarray}}
\newcommand{\eeqa}{\end{eqnarray}}
\newcommand{\eq}{\begin{equation}}
\newcommand{\eeq}{\end{equation}}

\newcommand{\kB}{k_{\rm B}}

\begin{document}
\title{Timescales for stochastic barrier crossing: inferring the potential from nonequilibrium data}

\author{A.~J.~Archer}

\affiliation{
 Department of Mathematical Sciences and Interdisciplinary Centre for Mathematical Modelling, Loughborough University, Loughborough, LE11 3TU, United Kingdom \\
}%
\author{T.~Ala-Nissila}

\affiliation{
 Department of Mathematical Sciences and Interdisciplinary Centre for Mathematical Modelling, Loughborough University, Loughborough, LE11 3TU, United Kingdom \\
}%
\affiliation{
 MSP group, Department of Applied Physics, Aalto University, P.O.~Box 15600, FI-00076 Aalto, Espoo, Finland. \\
}%

\author{T.~J.~W.~Honour}
\author{S.~P.~Fitzgerald}

\affiliation{
 Department of Applied Mathematics,
 University of Leeds, Leeds, LS2 9JT, United Kingdom \\
}%

\date{\today}

\begin{abstract}
\noindent
Kramers' rate theory forms a cornerstone for thermally activated barrier crossing. However, its reliance on equilibrium quantities excludes analysis of nonequilibrium dynamics at early times. Most works have thus focused on obtaining rates and transition time and path distributions in equilibrium. Instead, here we consider early-time nonequilibrium dynamics in a model system of a particle with overdamped dynamics hopping over the barrier in a double-well potential, using the Smoluchowski equation (SE) and stochastic path integral (SPI) mapping of the Langevin equation. We identify several key timescales relevant to nonequilibrium dynamics and quantify them using the SE and SPI approaches. The shortest timescale corresponds to equilibration in a well at time $t \ll \tau_{\rm B}$, where $\tau_{\rm B}$ is the Brownian diffusion time. The second important timescale is when an inflexion point appears in the effective potential constructed from the density at $t \lessapprox \tau_{\rm B}$. Shortly after, the existence of a second potential well can be inferred from sufficient sampling of the dynamics. Interestingly, this timescale decreases with increasing barrier height. We find significant deviations from the equilibrium limit unless $t \gg \tau_{\rm B}$. We further calculate the density current at the barrier for bistable and asymmetric potentials and find that it crosses over to that from equilibrium rate theory at a time that does not appear to depend on the barrier height. Our results have important implications for controlling activated processes at finite times and demonstrate the importance of reaching long enough times to faithfully construct potential landscapes from experimental or simulation data.
\end{abstract}

\maketitle

\section{Introduction}
Thermally activated processes and classical barrier crossing in particular are fundamental and ubiquitous processes whose equilibrium limit can be in many cases be understood by rate theories such as that by Kramers. 
Examples
include diffusion in periodic potentials \cite{TAN2002,Antczak_Ehrlich_2010}, escape from a metastable well or a local minimum \cite{hanggi1990reaction}, particles hopping in a bistable potential well \cite{ginot2022barrier, PhysRevResearch.6.043237, PhysRevX.14.011017}, particles in electromagnetic traps \cite{YoonSongLeeKimLeeKim+2020+4729+4735,PhysRevLett.125.146001}, chemical reactions \cite{srivastava2002encyclopedia}, protein folding \cite{C1CP21541H}, buckling transition in graphene \cite{PhysRevMaterials.8.016001}, etc.
At temperatures low compared to the barrier height, the timescale for a transition to occur becomes much longer than the relaxation time around the minimum.
For such rare events, it is often a good approximation to use quasi-equilibrium or linear response arguments to derive explicit formulae for the relevant physical quantities.
Kramers' escape rate formula relates the escape rate $r_{\rm e}$ to the exponential Arrhenius form for the normalized barrier height $e^{-\Delta F/(k_{\rm B}T)}$, where $\Delta F$ is the (free) energy barrier, $T$ is temperature, and $k_{\rm B}$ the Boltzmann constant.

In most cases of interest to date, the focus on barrier-crossing transitions has been in finding the rates and analysing the transition paths (TP).
In stochastic dynamics, TPs in an effective free-energy landscape or in an external (adiabatic) potential {are defined as paths that connect the initial state of the system to the final state in the reaction coordinate space \cite{bolhuis2002transition}.
In particular, the TP time distributions} can be used to obtain information about the barriers, Kramers' rates and even non-Markovian effects in systems with memory correlations \cite{PhysRevResearch.6.043237}.

With enough {experimental or simulation spatial trajectory statistics, it is possible to use this data} to reconstruct the effective free-energy landscape $F_{\rm eff}(\vec x)$ from the probability density distribution function of the particle position $\rho(\vec x)$ as \cite{makarov2021barrier, best2005reaction}
\eq 
    F_{\rm eff}(\vec x) = -k_{\rm B} T \ln \rho(\vec x).
    \label{eq:1}
\eeq 
This has been shown to work for symmetric bistable potentials in many different systems.  
However, a necessary condition for obtaining the true $F_{\rm eff}(\vec x)$ is that the probability density corresponds to the true asymptotic equilibrium Boltzmann density $\rho_{\rm B}(\vec x)$. Further, it it has recently been theoretically shown that {TP time distributions} cannot correctly capture {barriers and rates} from hopping in spatially anisotropic potentials \cite{caraglio2020transition}. 

Most recently, rare event theory has been extended to nonequilibrium processes with a focus on transitions between metastable nonequilibrium states \cite{singh2025reactivepathensemblesnonequilibrium, singh2025variational}.
However, even when dealing with near-equilibrium transitions, which are bound by principles such as detailed balance, there are nontrivial features in the dynamics during times less than the average transition time $t \ll \tau_{\rm e} = 1/r_{\rm e}$.
{For systems where the TP correspond to a rare event, the system spends most of its time stochastically sampling the reactively unproductive regions of configuration space.}
In higher-dimensional spaces, the system may explore saddle points and local minima which are not along the TPs.
{This has some potentially important implications for finite-time dynamics (i.e.\ dynamics sampled over a specified short time, such as in Ref.~\cite{fitzgerald2023stochastic}) and in particular to controlling chemical reactions and other activated processes before equilibration to the final state.
An important example of this is extraction and control of intermediates from slow chemical reactions. In many cases biochemical and biophysical reactions involve intermediate steps, where desirable reaction products appear and could be characterized and harvested. 
To facilitate this, it is necessary to understand and optimize reaction paths at finite times to achieve desired concentrations of intermediates \cite{doi:10.1021/acs.chemrev.2c00798, fitzgerald2023stochastic}.
A second example where early-time dynamics is important is protein folding, where the system explores the phase space before eventually finding its TP \cite{C1CP21541H}.}

The purpose of the present work is to theoretically study finite-time transition dynamics in a simple one-dimensional model of an overdamped particle in an external double-well potential.
To this end, we pursue a two-pronged approach based on (i) solving the Smoluchowski equation (SE) for the time evolution of the probability density of the particle, from which we can reconstruct the external potential landscape at finite times. We complement this with (ii) stochastic path integral (SPI) methods to determine the time taken after initiating the particle in one well, for the probability density to start building in the other well.

Our solutions of the SE show that while the density asymptotically approaches the expected Boltzmann distribution, we find that even at moderately low temperatures, there are still notable deviations at finite times from the equilibrium limit $t \gg \tau_{\rm B}$, where $\tau_{\rm B}$ is the diffusive Brownian time scale.
This has important implications for reconstructing the free energy landscape from the probability density $\rho(x,t)$.
Significant deviations from the true energy landscape occur up to times $t \approx \tau_{\rm B}$ and beyond.
We show how such deviations can be used as a resource to explore an unknown potential landscape. To this end, we identify several relevant timescales in the problem that play an important role in the dynamics.
The shortest is the equilibration time in a well at $t \ll \tau_{\rm B}$, which can be estimated from the SE.
The second timescale is when the effective potential constructed from $\rho(x,t)$ shows an inflexion point at a point $x$ in the vicinity of the second well, at time $t < \tau_{\rm B}$.
We compute this time numerically from the SE and show how it can be analytically obtained from the SPI.
The importance of this time is that it reveals the existence of another well for an unknown potential energy landscape.
{Interestingly}, this timescale decreases with increasing potential barrier.  
Further, we calculate the particle density current at the barrier from the SE and derive an analytic expression for the dependence of the current on the hopping rates for a bistable and asymmetric double-well potential.
We find that it crosses over to that predicted by equilibrium rate theory at a time that does not depend on the barrier height or asymmetry.

This paper is structured as follows:
We start in Sec.~\ref{sec:theory} with a brief introduction to the equations of motion of our system and the theories used in this study: SPI methods and the SE, which is the partial differential equation for the time evolution of the particle probability density $\rho(x,t)$.
In Sec.~\ref{sec:3} we present our results.
We start by showing in Sec.~\ref{subsec:3A} some typical results for $\rho(x,t)$ and then in Sec.~\ref{subsec:3B} explain and illustrate how these results lead directly to the effective potential that one would infer from particle trajectories sampled at time $t$.
We also calculate the {elapsed} time $t_{\rm d}$ for a point of inflexion to appear in the effective potential {after having been initiated in the left hand well,} and show how $t_{\rm d}$ depends on the size of the potential barrier.
Then, in Sec.~\ref{subsec:PI_results}, we present an alternative analytic approach based on SPI  for analytically calculating this time.
In Sec.~\ref{subsec:current}, we present results for the net particle current at the barrier over time, comparing the results from solving the SE with the results from a simple master equation together with the rates from equilibrium rate theory.
Finally, in Sec.~\ref{sec:4} we make a few concluding remarks.

\section{Theory}
\label{sec:theory}

\subsection{Equation of motion}

An overdamped noise-driven system in one dimension can be modelled by the Langevin equation \cite{risken, gardiner}
\eq 
\gamma \dot x(t) = -V'(x) + \xi(t),\label{eq:langevin}
\eeq  
where $x(t)$ is the coordinate, $V(x)$ is the external potential ($V'=dV/dx$), $\gamma$ is the friction coefficient, and $\xi$ is a Gaussian white noise of strength $\gamma\kB T$, where $\kB$ is Boltzmann constant and $T$ is the temperature.
Note that for higher-dimensional systems, the coordinate $x$ is the reaction coordinate (slow variable) and all other variables have been integrated out. In such cases, $V$ is actually the coarse-grained free energy $F_{\rm eff}$ in Eq.~\eqref{eq:1}.
The noise in Eq.~\eqref{eq:langevin} has correlation function
\eq 
\langle \xi(t)\xi(t')\rangle = 2\kB T\gamma\delta(t-t').
\eeq 
The fluctuation-dissipation theorem sets the diffusion coefficient to $D=\kB T/\gamma$.

The external potential that we consider here has the general form of a double-well potential,
\eq 
  V(x)=\epsilon\left(\frac{x+l}{l}\right)^2\left(\frac{x-l}{l}\right)^2-a\frac{x}{l}.
  \label{eq:V}
\eeq 
When the tilt parameter $a=0$, then this potential exhibits two minima at $x=\pm l$, with a local maximum (the barrier) at $x=0$, with barrier height $\Delta V=[V(0)-V(-l)]=\epsilon$.
When $a\neq0$, then the locations of the minima and the barrier maximum are slightly shifted.
Thus, we henceforth denote the location of the left hand minimum as  $x_{\rm L}$, the location of the right hand minimum as  $x_{\rm R}$ and the location of the barrier maximum as  $x_{\rm B}$.
Additionally, we use the length scale $l$ in $V(x)$ to scale all lengths and all times are given in units of the Brownian timescale $\tau_{\rm B}=l^2/D$.

\subsection{Stochastic path integral formalism}
\label{subsec:PI_formalism}

A powerful approach to stochastic processes, such as those described by the overdamped Langevin equation \eqref{eq:langevin}, is given by the path integral.
The noise $\xi$ in Eq.~\eqref{eq:langevin} has Gaussian probability distribution functional 
\eq 
\calP[\xi(t)]\propto \exp\left( -\frac1{4D\gamma^2}\int \xi^2\dee t\right).
\eeq 
Substituting Eq.~\eqref{eq:langevin} into this gives
\eq 
\calP[x(t)]\propto \exp\left( -\frac1{4\kB T\gamma}\int \left(\gamma\dot x + V'\right)^2\dee t\right),
\eeq 
for the weight attached to a system trajectory, and the transition probability can be written as an integral over such trajectories satisfying $x(t_0)=x_0; x(t_1)=x_1$ \cite{hunt1981path, adib2008stochastic, wio2013path}:
\eq 
P(x_1,t_1|x_0,t_0)=\int\calD x\,\calJ[x]\exp\left( -\frac1{4\kB T\gamma}\int \left(\gamma\dot x + V'\right)^2\dee t\right).
\label{eq:PI_withJ}
\eeq 
$\calJ$ is a Jacobian arising from changing variables in the functional integral from noise realizations $\xi$ to system trajectories $x$ (a path-independent constant is absorbed in the measure). It is given by (see, e.g., Refs.~\cite{hunt1981path, adib2008stochastic, wio2013path} and the Appendix)
\eq 
\calJ[x] = \exp\left(\frac{1}{2\gamma}\int_{t_0}^{t_1}V''(x(t))\dee t\right).
\label{eq:Jacobian}
\eeq
Thus, the transition probability can be written as
\eq 
P(x_1,t_1|x_0,t_0)=\int\calD x\,\exp\left(-\frac{\calS[x]}{4\kB T\gamma}\right),
\label{eq:PI}
\eeq 
where the stochastic (Onsager-Machlup) action $\calS$ is 
\eqa 
\calS[x] &=& \int_{t_0}^{t_1}\left[\left(\gamma\dot x + V'\right)^2-2\kB T V''\right]\dee t \nonumber \\
&=& 
2\gamma\left[V(x_1)-V(x_0)\right] +  \gamma\int_{t_0}^{t_1}\left(\gamma\dot x^2 + \frac{V'^2}{\gamma}-2D V''\right)\dee t.
\label{eq:action}
\eeqa 
The action splits into a path-independent part $2\gamma\Delta\!V = 2\gamma\left[V(x_1)-V(x_0)\right]
$ and a time-reversal-invariant dissipative remainder.
{As long as the barrier height $> k_{\rm B}T$}, the integral in Eq.~\eqref{eq:PI} can be evaluated in the saddle-point approximation, retaining only paths close to action-minimizing trajectories.
These extremal trajectories satisfy $\delta\calS/\delta x = 0$, and correspond to the Hamiltonian mechanics of a particle of mass $2\gamma^2$ moving in the potential
\eq
W(x) \equiv -\frac{V'(x)^2}{\gamma}+2D V''(x),
\label{eq:W}
\eeq
which we refer to as the `action potential'.
\begin{figure}[t!]
\centering
\includegraphics[width=0.6\textwidth]{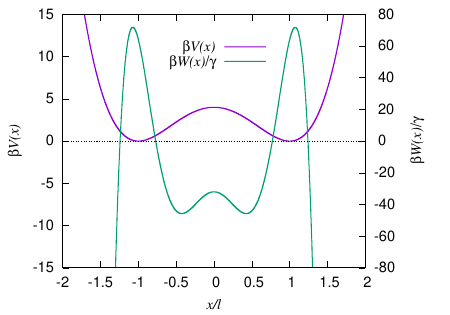}
\caption{Potential $V(x)$ for the case when the barrier height $\beta\epsilon=4$ and no tilt, $a=0$ (left axis scale). We also display (right axis scale) the action potential $W(x)$ for the Hamiltonian dynamics in Eq.~\eqref{eq:W}.}
\label{fig:pots}
\end{figure}
The smooth Hamiltonian trajectories correspond to the centre lines of tubes linking the initial and final points $(x_0,t_0)$ and $(x_1,t_1)$. These tubes (which get narrower as the noise weakens \cite{strat1971}) contain the most probable (non-smooth) stochastic trajectories that realize the transition. 

For the double-well potential \eqref{eq:V} considered here, we display in Fig.~\ref{fig:pots} both the potential $V(x)$ and also the corresponding action potential $W(x)$.
Note that in Fig.~\ref{fig:pots} we plot the dimensionless potentials $\beta V(x)$ and $\beta W(x)/\gamma$, where $\beta=(\kB T)^{-1}$.
The effective Hamiltonian mechanics has a conserved quantity $H$, given by 
\eq 
H = \gamma \dot x^2 +W(x). \label{eq:H}
\eeq 
This is the energy of the effective dynamics, but has the dimensions of power. $H/4D$ can be identified with the Laplace parameter of the Laplace-transformed dynamics \cite{fitzgerald2024}. Inserting this into $\calS$ gives the action evaluated along the extremal (Hamilton's principal function, $S(x_1,x_0,t_1-t_0)$)
\eq 
S = 2\gamma\Delta\! V - \gamma H(t_1-t_0) + 2\gamma\int_{\Gamma}\sqrt{\gamma(H-W)}\dee s,
\eeq 
where the integral is evaluated along the path $\Gamma$.
In one dimension, $\dee s = \pm\dee x$.
We also use the result from Eq.~\eqref{eq:H} that $\gamma \dee s=\sqrt{\gamma(H-W)}\dee t$.
An equation for $H$ can be derived by integrating Eq.~\eqref{eq:H} (or equivalently by setting $\partial S/\partial H = 0$), to obtain:
\eq 
t_1-t_0 = \sqrt{\gamma}\int_{\Gamma}\frac{\dee s}{\sqrt{H-W}}.
\label{eq:time}
\eeq 
This gives the time associated with the path $\Gamma$.
We use this result below in Sec.~\ref{subsec:PI_results} to estimate the time for the particle probability density to start building up in one well, with the particle having been initiated at time $t_0=0$ in the other well.

\subsection{Dynamics of the probability density}
\label{subsec:Smoluchowski}

A complementary approach for understanding the dynamics of a particle with the stochastic equation of motion \eqref{eq:langevin}, is via the time evolution of the probability density $\rho(x,t)$.
This is given by the following Smoluchowski {(overdamped Fokker-Planck)} equation \cite{risken}:
\eq 
\frac{\partial \rho}{\partial t}=\frac{1}{\gamma}\frac{\partial }{\partial x}\left(\kB T\frac{\partial \rho}{\partial x}+\rho\frac{\partial V}{\partial x}\right).
\label{eq:Smoluch}
\eeq 
This dynamical equation has the form of the continuity equation:
\eq 
\frac{\partial \rho}{\partial t}=-\frac{\partial j}{\partial x},
\label{eq:continuity}
\eeq 
where the current
\eq 
j(x,t)=-D\frac{\partial \rho}{\partial x}-\frac{\rho}{\gamma}\frac{\partial V}{\partial x}.
\label{eq:current}
\eeq 
We solve the SE~\eqref{eq:Smoluch} numerically using the finite difference algorithm described in Ref.~\cite{chalmers2017dynamical}, with spatial grid spacing $\Delta x/l=0.015$ and time step $\Delta t /\tau_{\rm B}=10^{-6}$.
If we consider starting the particle in the left hand well at $x=x_{\rm L}$ at time $t=0$, then the initial profile is $\rho(x,t=0)=\delta(x-x_{\rm L})$, where $\delta(x)$ is the Dirac $\delta$ distribution.
Putting this as the initial condition into a finite-difference numerical solver is of course not possible.
Therefore, we assume the analytic free-diffusion (i.e.\ $V=0$) solution
\eq 
    \rho_{0}(x,t)=\frac{e^{-(x-x_{\rm L})^2/4Dt}}{\sqrt{4\pi Dt}}
    \label{eq:free_diff}
\eeq 
applies at very early times, and use Eq.~\eqref{eq:free_diff} at the time $t_{\rm i}=0.005\tau_{\rm B}$ as our initial condition.

\section{Results}
\label{sec:3}

\subsection{Solution of the Smoluchowski equation}
\label{subsec:3A}

\begin{figure}
    \centering
    \includegraphics[width=0.7\linewidth]{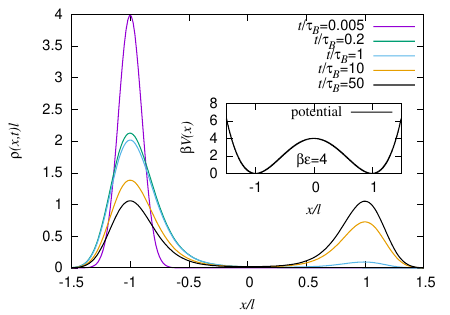}
    \caption{The probability density $\rho(x,t)$ at various times (given in the key), corresponding to a particle starting at $t=0$ at the minimum $x=x_{\rm L}=-l$ of the left hand well of the potential $V(x)$ displayed in the inset. The height of the potential barrier $\beta\epsilon=4$ and in this case there is no tilt to the potential, $a=0$. We see that the density initially spreads throughout the left-hand well and then over time, the likelihood of the particle having moved to the right-hand well increases. By the time $t=50\tau_{\rm B}$, the probability of the particle being in either of the two wells is equally likely.}
    \label{fig:rhos}
\end{figure}

In Fig.~\ref{fig:rhos} we present some typical results from solving the SE for the probability density $\rho(x,t)$, corresponding to the case when $a=0$ (no tilt) and barrier height $\beta\epsilon=4$.
We see that over a short time $\approx0.2\tau_{\rm B}$ the probability density in the interval $-1.6\lesssim x/l\lesssim-0.3$ around the left-hand minimum is far from zero, corresponding to the particle exploring all of the left-hand potential well.
By the time $t\approx \tau_{\rm B}$ the beginnings of a probability density peak in the right hand well is noticeable, indicating the chance of the particle having crossed the barrier in this time is non-negligible.
In the long time limit, the equilibrium probability density distribution is the Boltzmann distribution
\eq 
\rho_{\rm B}(x,t\to\infty)=\rho_{\rm eq}e^{-\beta V(x)},
    \label{eq:equilibrium}
\eeq 
where the constant $\rho_{\rm eq}$ is determined by the normalization $\int_{-\infty}^\infty\rho_{\rm B}(x,t\to\infty)\dee x=1$.
When $a=0$, this distribution is an even function, with 50\% of the probability to the left of the barrier at $x_{\rm B}=0$ and the remaining 50\% to the right.
Figure~\ref{fig:rhos} shows that by the time $t\approx50\tau_{\rm B}$ the distribution $\rho(x,t)$ is almost indistinguishable by visual inspection from the equilibrium profile of Eq.~\eqref{eq:equilibrium}.
{Notice too} that even after the relatively long time $t=10\tau_{\rm B}$, the system is far from equilibrated.
This illustrates the fact that it is unlikely but possible to observe a transition over the barrier at fairly early times after placing the particle at $x_{\rm L}$ in the left hand well, but equilibration takes a much longer time.
Reaction rates are obtained in the long-time limit, so this illustrates the fact that one should be cautious about estimating rates from data gathered after a period less than the equilibration time.
{Of course, it is desirable to simulate for shorter periods, and many computational techniques have been developed to this end \cite{tuckerman2023statistical}.}
In Sec.~\ref{subsec:current} we show how to determine this equilibration time.
However, before doing that, we consider how results (e.g., from experiments) for $\rho(x,t)$ can be used to infer the shape of the potential~$V(x)$.

\subsection{Inferring the potential from nonequilibrium data}
\label{subsec:3B}

\begin{figure}
    \centering
    \includegraphics[width=0.7\linewidth]{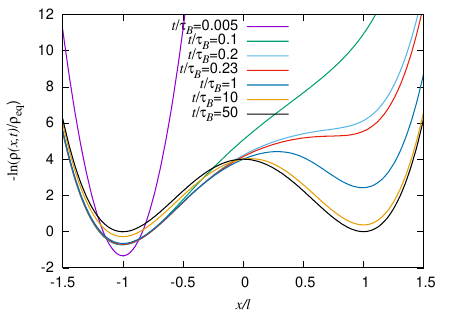}
    \caption{Plot of the scaled effective potential $\beta V_{\rm eff}(x,t) = -\ln(\rho(x,t)/\rho_{\rm eq})$ over time. In the limit $t\to\infty$, this gives the true external potential $\beta V(x)$ [cf.\ Eq.~\eqref{eq:equilibrium}] for the case when the barrier height $\beta\epsilon=4$ and tilt $a=0$. We see that at the time $t/\tau_{\rm B}\approx0.23$ a point of inflexion appears in the profile near $x\approx x_{\rm R}$ and that for times greater than this, there is then a second minimum in $V_{\rm eff}$.}
    \label{fig:log_rho}
\end{figure}

Suppose we sample the trajectory of a Brownian particle (e.g., the experimental data presented in Ref.~\onlinecite{PhysRevX.14.011017}), and we wish to determine the potential $V(x)$ in which the particle is moving.
Rearranging Eq.~\eqref{eq:equilibrium}, we can write the potential in terms of the $t\to\infty$ equilibrium density distribution as $V(x)=-\kB T\ln(\rho/\rho_{\rm eq})$.
The obvious question of practical importance is: what finite time is sufficient for this result to be used and still be accurate?
To address this question, in Fig.~\ref{fig:log_rho} we display plots of the time-dependent effective potential $\beta V_{\rm eff}(x,t) = -\ln(\rho(x,t)/\rho_{\rm eq})$ over time $t$ for the same situation as the results presented in Fig.~\ref{fig:rhos}.
As already foreshadowed in our discussion of Fig.~\ref{fig:rhos}, to reconstruct the true potential $V(x)$, we see from Fig.~\ref{fig:log_rho} that one needs to wait until times $t\approx50\tau_{\rm B}$ or later. This highlights the importance of having to reach the long-time limit $t \gg \tau_{\rm B}$ to properly reconstruct the potential landscape.

A particularly interesting feature in $V_{\rm eff}$ is the appearance of an inflexion point at the time $t_{\rm d}=0.233\tau_{\rm B}$ in the effective potential near $x_{\rm R}$.
The significance of this time $t_{\rm d}$ is that if one samples the density distribution and the effective potential before this time, there is no indication of a second minimum in the potential.
However, if one samples after this time, then there is a second minimum in the effective potential $V_{\rm eff}$, or equivalently a second maximum in the density distribution $\rho(x,t)$. 
That there is a point $x\approx x_{\rm B}$ to the right of the starting point $x_{\rm L}$ where the density is lower than a point even further to the right at $x\approx x_{\rm R}$ is enough data to infer there is indeed a second minimum to the potential.
Of course, this time is not long enough to determine the depth of the second minimum, but the short time $t>t_{\rm d}$ is long enough to infer its existence. In Fig.~\ref{fig:timescales} we show a plot of the value of $t_{\rm d}$ as a function of the barrier height $\beta\epsilon$ for situations similar to that in Fig.~\ref{fig:log_rho}, where $a=0$.
These values of $t_{\rm d}$ are calculated by solving the SE and identifying the time when a point of inflexion appears in $V_{\rm eff}(x)$ near the point $x_{\rm R}$.

\subsection{Timescale from the stochastic path integral}
\label{subsec:PI_results}

\begin{figure}
    \centering
    \includegraphics[width=0.7\linewidth]{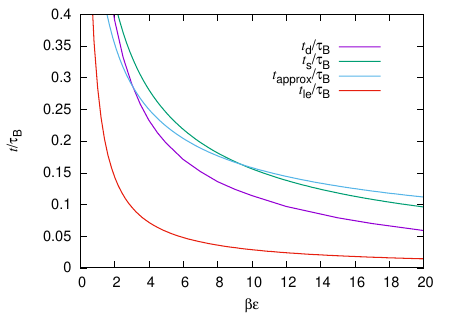}
    \caption{A plot of some of the relevant times as a function of the barrier height, $\beta\epsilon$. The time $t_{\rm d}$ is the time taken for the second minimum at $x\approx x_{\rm R}$ to appear in the effective potential $V_{\rm eff}(x,t)$. The results for $t_{\rm d}$ are obtained from numerical solution of the SE. We also plot the time $t_{\rm s}$, which is our estimate for this time obtained using the stochastic path integral formalism, together with the very simple analytic approximation $t_{\rm approx}$ in Eq.~\eqref{eq:t_approx} for these times. The very different time also plotted, $t_{\rm le}$, is the time taken for a particle to locally equilibrate in the left well. This time is much shorter than the other (barrier crossing) timescales, except in the limit $\beta\epsilon\to0$ (not considered here).}
    \label{fig:timescales}
\end{figure}

We now go back to the SPI formalism outlined in Sec.~\ref{subsec:PI_formalism} and show how this theory can be used to obtain a timescale $t_{\rm s}$ which has a very similar value to $t_{\rm d}$, the time for a point of inflexion to appear in $V_{\rm eff}(x)$ near $x_{\rm R}$. 
For a second peak to appear in the well near $x_{\rm R}$, $V_{\rm eff}$ must have a local minimum there. {This means} that the action is a decreasing function of $x$ for some interval ending at $x_{\rm R}$, and hence that the Hamiltonian trajectory approaches its endpoint from the right. In other words, the particle must leave its initial point at $x_{\rm L}$, pass through its eventual endpoint, then turn around and return. This is only possible for values of $H$ less than the maximum of $W$ near $x_{\rm R}$, since trajectories with higher energies would surmount the maximum of $W$ and continue to higher values of $x$. The limiting case, corresponding to the emergence of the inflexion point in $V_{\rm eff}$, is when the particle leaves $x_{\rm L}$ with zero kinetic energy and comes to rest at $x_{\rm R}$. This trajectory will have a finite time, since the maxima of $W$ are offset from $x_{\rm L}$ and $x_{\rm R}$, the minima of $V$ (see Fig.~\ref{fig:pots}).

The integral in Eq.~\eqref{eq:time} gives the time corresponding to the extremal path with power $H$.
We now use this result to calculate the time associated with a path starting in the left hand well at $x_0=x_{\rm L}$ at time $t_0=0$.
We assume that the initial kinetic energy is zero, and therefore Eq.~\eqref{eq:H} gives the value of the power $H$ for this path to be
\eq
    H_{\rm s}\equiv W(x_{\rm L})=2DV''(x_{\rm L}),
    \label{eq:H_s}
\eeq
where we have used Eq.~\eqref{eq:W} and the fact that our starting point is a local minimum in $V$, to obtain the second expression above for $H=H_{\rm s}$.
Taking this path to terminate in the right-hand well at $x_{\rm R}$, and then putting these limits and specific value of $H$ into Eq.~\eqref{eq:time}, we obtain
\eq
    t_{\rm s} = \sqrt{\gamma}\int_{x_{\rm L}}^{x_{\rm R}}\frac{\dee x}{\sqrt{H_{\rm s}-W(x)}},
    \label{eq:t_s}
\eeq
with $W(x)$ given in Eq.~\eqref{eq:W}.
The value of $H=H_{\rm s}$ is the largest value of $H$ for which the path turns around in the Hamiltonian dynamics{, because the kinetic energy cannot overcome the potential barrier}.
The results for $t_{\rm s}$ from evaluating the integral in Eq.~\eqref{eq:t_s} for a range of different values of the barrier height $\epsilon$ are displayed in Fig.~\ref{fig:timescales}.
This figure shows that $t_{\rm d}\approx t_{\rm s}$, i.e., that these timescales are very similar.
They are not exactly the same, but given that many of the other relevant timescales in the problem, such as the overall equilibration time (see below), are much longer times compared to the timescales $t_{\rm b}$ and $t_{\rm s}$, we conclude that these two times are related to the same event, i.e.\ when barrier crossing becomes statistically noticeable.
A very simple estimate $t_{\rm approx}\approx t_{\rm s}$ can be obtained by setting $W=0$ in Eq.~\eqref{eq:t_s}{.
This gives}
\eq
   t_{\rm approx}=2l\sqrt{\frac{\gamma}{H_{\rm s}}}.
   \label{eq:t_approx}
\eeq
A plot of this time as a function of $\beta\epsilon$ is also displayed in Fig.~\ref{fig:timescales}, which shows that the three timescales $t_{\rm d}$, $t_{\rm s}$ and $t_{\rm approx}$ are similar.
That said, there are some small differences between the dependence on the barrier height of these three times.
On a double logarithmic plot of these times as a function of $\beta\epsilon$ (not displayed), only $t_{\rm approx}$ shows exactly straight-line behaviour.
This is easy to see from Eqs.~\eqref{eq:H_s} and \eqref{eq:t_approx}, which together yield that $t_{\rm approx}\sim\epsilon^{-1/2}$.
Even over the range of $\beta\epsilon$ values for which we have calculated the times $t_{\rm d}$ and $t_{\rm s}$, both of these times show deviations from this power-law behaviour {(which can also be inferred from Fig.~\ref{fig:timescales})}.

In Fig.~\ref{fig:timescales}, we also display a plot of the local equilibration time $t_{\rm le}$, which is the time it takes a particle to locally equilibrate in the left hand well.
This time $t_{\rm le}$ is substantially shorter than the three timescales discussed above, which are all related to crossing the barrier.
The time $t_{\rm le}$ can be estimated for sufficiently high barriers (or {low temperatures}) as follows:
This local equilibrium corresponds to a narrow density distribution centred on $x_{\rm L}$, that spreads only a relatively short distance from the initial $\delta$-distribution.
This can be approximately quantified with an Ornstein-Uhlenbeck approximation for the density in the left well.
The exact SE solution for the single-well potential $V(x) = \mu x^2/2$ (starting at $x=0$ when $t=0$) is {\cite{risken}}
\eq 
\rho(x,t) = \sqrt{\frac{\mu}{2\pi D\gamma(1-e^{-2\mu t/\gamma})}}\;\exp\left( -\frac{\mu x^2}{2D\gamma(1-e^{-2\mu t/\gamma})} \right).
\eeq 
The time taken to equilibrate is controlled by the term $(1-e^{-2\mu t/\gamma})$ which depends only on the ratio $\mu/\gamma$.
Using this to estimate the local equilibration time in the left well of the double-well potential \eqref{eq:V}, this translates to $\mu = V''(-l) = 8\epsilon/l^2$.
Taking local equilibration as being when $e^{-2\mu t/\gamma}<0.01$ (i.e., ``1\% away from equilibrium''), gives the time
\eq
t_{\rm le}=\frac{\alpha\gamma}{2V''(x_{\rm L})},
\label{eq:t_le}
\eeq
where $\alpha=-\ln(1\%)\approx4.6$.
For example, for the case when $\beta\epsilon=4$, the above gives {$t_{\rm le}\approx 0.07\tau_{\rm B}$}, which is shorter than all of the timescales discussed so far.
Since $V''(x_{\rm L})\sim\epsilon$, Eq.~\eqref{eq:t_le} shows that the local equilibration time scales as $t_{\rm le}\sim\epsilon^{-1}$, which is very different from the behaviour of $t_{\rm d}$ and $t_{\rm s}\approx t_{\rm approx}\sim\epsilon^{-1/2}$ discussed previously.
This estimate breaks down when either the {barrier is too small} (too much density crosses the barrier before local equilibration, i.e., the timescales are insufficiently separated) or the quadratic approximation to the well is not accurate (e.g.\ for a flat-bottomed well).

\subsection{Current at the barrier and the long-time limit}
\label{subsec:current}

Having considered in the previous subsection the time it takes for a particle to locally equilibrate in the left well $t_{\rm le}$ and also the time $t_{\rm d}$ at which barrier crossing can first be identified, we now consider the time it takes the system to fully equilibrate.
This occurs when the forward and backward rates for the transitions across the barrier are equal, or equivalently when the net flux (probability current) across the barrier is zero.
We calculate this via numerical solution to the SE, as described in Sec.~\ref{subsec:Smoluchowski}.
The maximum of the barrier, where $V'(x)=0$, is at $x=x_{\rm B}$.
At this point, the current \eqref{eq:current} is
\eq 
j(x=x_{\rm B})=-D\frac{\partial \rho}{\partial x}\Big|_{x=x_{\rm B}}.
\eeq 
Integrating Eq.~\eqref{eq:continuity} with respect to $x$ gives
\eq 
\frac{\partial }{\partial t}\int_{-\infty}^{x_{\rm B}}\rho \dee x
=-\int_{-\infty}^{x_{\rm B}}\frac{\partial j}{\partial x}\dee x,
\eeq 
and from this we obtain
\eq 
    \frac{\partial P_{\rm L}}{\partial t}=-\left[j(x,t)\right]^{x_{\rm B}}_{-\infty} = -j(x={x_{\rm B}},t),
    \label{eq:j_0_formula}
\eeq 
where $P_{\rm L}(t)\equiv\int_{-\infty}^{x_{\rm B}}\rho(x,t) \dee x$ is the probability of finding the particle in the left hand well. Due to the conservation of probability, the probability of being in the right hand well $P_{\rm R}(t)\equiv\int_{x_{\rm B}}^{\infty}\rho \dee x=1-P_{\rm L}(t)$.

The dynamics of the probability function can also be described by the master equation \cite{hanggi1990reaction}
\eqa
    \frac{\partial P_{\rm L}}{\partial t}&=&r_{{\rm L}\to {\rm R}}P_{\rm L}-r_{{\rm R}\to {\rm L}}P_{\rm R}\nonumber \\
    &=&r_{\rm L\to R}P_{\rm L}-r_{\rm R\to L}(1-P_{\rm L}),
    \label{eq:master}
\eeqa
where $r_{{\rm L}\to{\rm R}}$ and $r_{{\rm R}\to {\rm L}}$ are the rates for hopping from the left well to the right well and the reverse, respectively.
If we assume that the rates $r_{{\rm L}\to {\rm R}}$ and $r_{{\rm R}\to {\rm L}}$ are constants, then Eq.~\eqref{eq:master} can be integrated to obtain
\eq 
    P_{\rm L}(t)=[P_{\rm L}(0)-P_{\rm L}(\infty)]e^{-(r_{\rm L\to R}+r_{\rm R\to L})t}+P_{\rm L}(\infty).
    \label{eq:P_L}
\eeq 
Inserting the above result into Eq.~\eqref{eq:j_0_formula}, we obtain the following expression for the current at the barrier:
\eq 
    j(x=x_{\rm B},t)=[1-P_{\rm L}(\infty)](r_{\rm L\to R}+r_{\rm R\to L})e^{-(r_{\rm L\to R}+r_{\rm R\to L})t},
    \label{eq:j_0_general}
\eeq 
where we have used our initial condition that the particle starts in the left well, i.e., that $P_{\rm L}(0)=1$.

In the long-time limit, the above rates $r_{\rm L\to R}$ and $r_{\rm R\to L}$ correspond to the reciprocals of the mean first passage times, which can be calculated by integrating the SE~\eqref{eq:Smoluch}, assuming a steady-state with boundary conditions corresponding to inserting probability density in the left well at the same rate as removing it from the right well, to obtain \cite{risken}
\eq 
    r_{\rm L\to R}^{\rm M}=\left[\frac{2}{D}\int_{x_{\rm L}}^{x_{\rm B}}e^{\beta V(x')}\int_{-\infty}^{x'}e^{-\beta V(x)}{\rm d}x {\rm d}x'\right]^{-1}.
    \label{eq:rate}
\eeq 
We note that this formula can often be approximated by the Kramers' rate equation as
\eq 
    r_{\rm L\to R}^{\rm K}=\frac{\sqrt{|V''(x_{\rm L})V''(x_{\rm B})|}}{2\pi}e^{-\beta [V(x_{\rm B})-V(x_{\rm L})]}.
    \label{eq:Kramers_rate}
\eeq 
The above expressions assume that the system is in a steady-state equilibrium, which is strictly valid only in the limit $t \to \infty$, due to the way we initialise the system.
In what follows, we use Eq.~\eqref{eq:rate} to calculate the rates $r_{\rm L\to R}$ and $r_{\rm R\to L}$ in Eq.~\eqref{eq:j_0_general}, in order to compare with the results from numerical solution of the SE.

\subsubsection{Equilibration in the zero tilt $a=0$ case}

We consider first the case of the spatially symmetric potential where there is no tilt, $a=0$.
We start the particle at $t=0$ in the left well at $x_{\rm L}=-l$, with probability of being in the left hand well $P_{\rm L}(t=0)=1$.
Since $a=0$, in the long time limit $t\to\infty$, $P_{\rm L}(\infty)= P_{\rm R}(\infty)=1/2$.
Moreover, the two rates are equal, $r_{\rm L\to R}=r_{\rm R\to L}\equiv r$.
Equation \eqref{eq:P_L} then gives
$ P_{\rm L}(t)=(e^{-2rt}+1)/2$.
Inserting this into Eq.~\eqref{eq:j_0_formula}, we obtain that the current at the barrier is the very simple expression, 
$j(x=x_{\rm B},t)=re^{-2rt}$.

\begin{figure}
    \centering
    \includegraphics[width=0.7\textwidth]{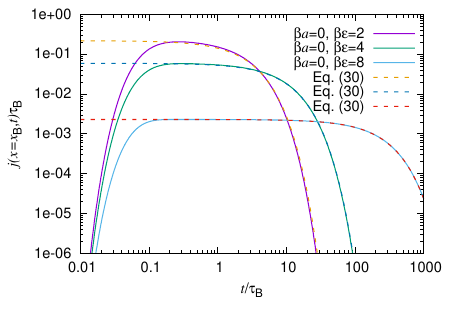}
    \caption{The current at the barrier as a function of time for the $a=0$ case (no tilt), in a double log plot. For three different values of the barrier height $\epsilon$ (given in the key) we display both the results from solving the exact SE (solid lines) and the asymptotic result in Eq.~\eqref{eq:j_0_general} (dashed lines). We see that for times $t\gtrsim\tau_{\rm B}$ these are in good agreement.}
    \label{fig:j_0}
\end{figure}

In Fig.~\ref{fig:j_0}, we present results for the current at the barrier $j(x=x_{\rm B},t)$ for three different values of the barrier height, obtained from numerically integrating the SE \eqref{eq:Smoluch}, as described above in Sec.~\ref{subsec:Smoluchowski}.
The data show a rapid increase in the flux at early times $t \ll \tau_{\rm B}$ corresponding to the initial spreading of the density.
This is followed by a rapid crossover to exponential decay, as given by Eq.~\eqref{eq:j_0_general}. 
In Fig.~\ref{fig:j_0}, we also plot the result in Eq.~\eqref{eq:j_0_general} using the mean first passage rates $r_{\rm R\to L}^{\rm M}$ and $r_{\rm L\to R}^{\rm M}$ obtained from Eq.~\eqref{eq:rate}, using the dashed lines.
The exponential decay law is well satisfied at late times.
Interestingly, we observe that the crossover time $t_{\rm c}$ from the initial rise to the exponential decay does not seem to depend on the barrier height (at least, not in an obvious way), with a typical value being roughly $t_{\rm c} \approx 0.2\tau_{\rm B}${, based on the data in Fig.~\ref{fig:j_0}}. For times $t>t_{\rm c}$, we see the exponential barrier-crossing law \eqref{eq:j_0_general}.
Note that this crossover time $t_{\rm c}$ is significantly larger than the time for local equilibration in the left well to occur, $t_{\rm le}$ in Eq.~\eqref{eq:t_le}; see also Fig.~\ref{fig:timescales}.

The expression for the flux at the barrier in Eq.~\eqref{eq:j_0_general}, valid for $t>t_{\rm c}$, can also be used to obtain an estimate for the time $t_{\rm eq}$ that the system takes to equilibrate, i.e., for the net current at the barrier to become $\approx 0$.
When $a=0$, this is the time it takes for the system to have roughly equal probability density in each well.
The equilibration time $t_{\rm eq}$ can be defined as the time when the current has decreased to a small fraction (here chosen to be 1\%) of is value at time $t=\tau_{\rm B}>t_{\rm c}$.
This gives
\eq 
    \frac{j(x=x_{\rm B},t=t_{\rm eq})}{j(x=x_{\rm B},t=\tau_{\rm B})}=\frac{[1-P_{\rm L}(\infty)](r_{\rm L\to R}+r_{\rm R\to L})e^{-(r_{\rm L\to R}+r_{\rm R\to L})t_{\rm eq}}}{[1-P_{\rm L}(\infty)](r_{\rm L\to R}+r_{\rm R\to L})e^{-(r_{\rm L\to R}+r_{\rm R\to L})\tau_{\rm B}}}=\frac{e^{-(r_{\rm L\to R}+r_{\rm R\to L})t_{\rm eq}}}{e^{-(r_{\rm L\to R}+r_{\rm R\to L})\tau_{\rm B}}}=1\%.
\eeq 
This can then be rearranged to give
\eq 
    t_{\rm eq}=\tau_{\rm B}+\frac{\alpha}{r_{\rm L\to R}+r_{\rm R\to L}},
    \label{eq:t_eq}
\eeq 
where again $\alpha=-\ln(1\%)\approx4.6$. For the three cases in Fig.~\ref{fig:j_0}, where the barrier heights considered are $\beta\epsilon=2$, 4 and 8, Eq.~\eqref{eq:t_eq} gives the times $t_{\rm eq}\approx12\tau_{\rm B}$, $40\tau_{\rm B}$ and $1008\tau_{\rm B}$, respectively. As 
Fig.~\ref{fig:j_0} shows, the results from the simple formula \eqref{eq:t_eq} predicts remarkably well the equilibration time.
When the barrier becomes high (larger $\beta\epsilon$), then the equilibration time \eqref{eq:t_eq} can be combined with the Kramers approximation for the rate \eqref{eq:Kramers_rate}, which together with Eq.~\eqref{eq:V} yields
\eq
t_{\rm eq}\approx\frac{\pi\alpha}{4\sqrt{2}\epsilon}\exp(\beta\epsilon).
\label{eq:t_eq_Kramers}
\eeq
This time is much greater than all three of the times $t_{\rm le}$, $t_{\rm d}$ and $t_{\rm c}$ and cannot be displayed on Fig.~\ref{fig:timescales}, without significantly changing the vertical time scale.

\subsubsection{Equilibration in the tilted $a\neq0$ potential}

\begin{figure}
    \centering
    \includegraphics[width=0.7\textwidth]{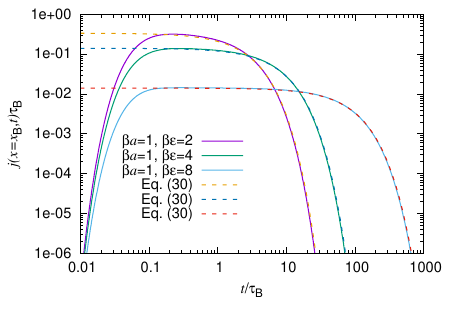}
    \caption{The current at the barrier as a function of time in a double log plot for various barrier heights $\beta\epsilon$, similar to Fig.~\ref{fig:j_0}, except here we display results for the tilted potential, with $\beta a=1$. The solid lines are the results from solving the SE and the dashed lines are the asymptotic result in Eq.~\eqref{eq:j_0_general}.}
    \label{fig:j_0_A1}
\end{figure}

\begin{figure}
    \centering
    \includegraphics[width=0.7\textwidth]{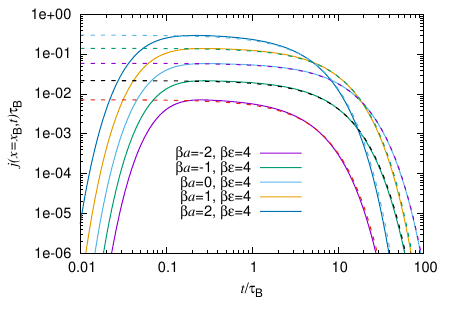}
    \caption{The current at the barrier as a function of time for varying values of the potential tilt parameter $a$ (given in the key), for fixed barrier height $\beta\epsilon=4$, in a double log plot. The solid lines are the results from solving the SE and the dashed lines are the corresponding asymptotic result in Eq.~\eqref{eq:j_0_general}.}
    \label{fig:j_0_beta1}
\end{figure}

Having considered the equilibration process for cases when there is no tilt to the potential $V(x)$, we now move on to briefly discuss cases where the tilt parameter $a\neq0$.
Figure~\ref{fig:j_0_A1} shows results for $\beta a=1$, which are very similar to the results in Fig.~\ref{fig:j_0} (which are for $a=0$),
except the final equilibration occurs very slightly faster.
The equilibration time in a tilted potential is not greatly different from that in the potential with no tilt because the equilibration time $t_{\rm eq}$ in Eq.~\eqref{eq:t_eq} depends on the sum of the forward and backwards rates, $(r_{\rm L\to R}+r_{\rm R\to L})$.
This means that when the potential $V(x)$ is tilted, the equilibration rate is the same whether the particle is started in the left well, or started in the right well.
This can be seen even more clearly in Fig.~\ref{fig:j_0_beta1}, where we show the current at the barrier $j(x=x_{\rm B},t)$, for fixed $\beta\epsilon=4$ and varying tilt parameter, $\beta a=\pm2$, $\beta a=\pm1$ and also $\beta a=0$, for comparison.
Figure~\ref{fig:j_0_beta1} is a double logarithmic plot, so small differences in the decay as $t\to\infty$ are highlighted.
That said, even on this scale, the final equilibration times are all rather similar.

\section{Summary and conclusions}
\label{sec:4}

In the present work, we have studied the ubiquitous activated barrier crossing problem from a point of view mostly neglected in the past, namely focusing on the early and finite-time nonequilibrium dynamics of the particle probability density.
To this end, we have used the Smoluchowski equation and stochastic path integral mapping of the overdamped Langevin equation as analytic and numerical tools.
Our system here is a one-dimensional double-well potential, where thermally activated hopping occurs starting from one of the wells.
As our main result, we find that there are multiple relevant timescales in the problem before approaching the equilibrium limit $t \gg \tau_{\rm B}$, where $\tau_{\rm B}$ is the diffusive Brownian time.
Specifically, we have identified and discussed four relevant timescales: $t_{\rm le}$, $t_{\rm d}$ (equivalently $t_{\rm s}$), $t_{\rm c}$ and $t_{\rm eq}$.
The shortest of these times, $t_{\rm le}$, is the time taken for the particle to reach local equilibrium in the left well of the potential.
This time scales as $t_{\rm le}/\tau_{\rm B}\sim\epsilon^{-1}$, where the parameter $\epsilon$ both determines the curvature of the potential at the minima and also determines the barrier height.
The next shortest time is $t_{\rm d}$ (equivalently $t_{\rm s}$) which both (roughly) scale as $t_{\rm d}/\tau_{\rm B}\sim\epsilon^{-1/2}$.
This is the time it takes for it to be possible to identify barrier crossing in the statistics of the particle density distribution $\rho(x,t)$.
The third time identified $t_{\rm c}$ is the time it takes for the current at the barrier to crossover to that predicted by equilibrium rate theory.
This time scales as $t_{\rm c}/\tau_{\rm B}\sim O(1)$ and does not seem to depend on the barrier height.
Finally, the longest timescale in the problem is the global equilibration time $t_{\rm eq}$, which is given very accurately by Eq.~\eqref{eq:t_eq} and equilibrium rate theory.
Using Kramers theory for the rate \eqref{eq:Kramers_rate}, we find that  $t_{\rm eq}\sim\exp(\beta\epsilon)/\epsilon$; see Eq.~\eqref{eq:t_eq_Kramers}.
This timescale is well understood \cite{hanggi1990reaction}, as is the manner in which this time grows with increasing barrier height $\epsilon$.
The {interesting} thing to come out of our study is that the time $t_{\rm d}$ in contrast gets shorter as the barrier gets higher.
This shows that in the statistics of rare events, the existence of a transition over a barrier can, with enough data, be identified after only sampling for a relatively short time.
However, if the height of the barrier and the depth of the potential in the target state (the right well in our model) are to be inferred, then a much longer sample time $t\approx t_{\rm eq}$ is required.

Despite the simplicity of our one-dimensional model, the work presented here should be useful to experimentalists and computer simulators who use time-sequence data for Brownian particles moving in an unknown potential in order to reconstruct the potential (free energy landscape) via Eq.~\eqref{eq:1} -- see, e.g., Ref.~\onlinecite{makarov2021barrier}
and also Ref.~\onlinecite{singh2024splitting}, where an approach to bias transitions is discussed.
Our results are also relevant to transition path sampling \cite{bolhuis2002transition, allen2009forward, keller2024dynamical}{, giving useful guidelines to the timescales needed for accurate path sampling, which should ideally extend to times much longer than the Brownian timescale}.
Our results show that if data is recorded over a time less than $t_{\rm eq}$ in Eq.~\eqref{eq:t_eq}, then the effective potential that is reconstructed is unreliable.
This is clearly illustrated in Fig.~\ref{fig:log_rho}, where the effective potential obtained after a seemingly long time $t=10\tau_{\rm B}$, is visibly different from the true potential $V(x)$.
This long time is significantly greater than the corresponding $t_{\rm d}=0.28\tau_{\rm B}$, the time associated with seeing barrier crossings in the particle density distribution.
This shows that observing multiple barrier crossings is not an indicator that the system has equilibrated and that the influence/bias from how the system is initiated takes a remarkably long time to drop out of the particle position statistics.

\section*{Acknowledgements}

We thank Sasha Balanov and Thomas Bartsch for valuable discussions. T.A-N. has been supported in part by the Academy of Finland grant no.~353298 under the European Union – NextGenerationEU instrument. S.P.F. was supported by the UK EPSRC, grant number EP/R005974/1. 

\appendix

\section{Stochastic path integral Jacobian}

The most straightforward way to calculate the functional Jacobian in Eq.~\eqref{eq:Jacobian} (see e.g. \cite{wio2013path}) is to discretize the Langevin equation \eqref{eq:langevin} over $N$ time steps of length $\Delta t$, and note that 
\eq 
\calJ = \left|\frac{\delta \xi}{\delta x} \right| = \lim_{N\to\infty} \det \frac{\dee \xi_i}{\dee x_j}.
\eeq
A discretization of Eq.~\eqref{eq:langevin} is
\eq 
\gamma(x_i - x_{i-1}) = -\left(\lambda\partial_x V(x_i) + (1-\lambda)\partial_x V(x_{i-1})\right)\Delta t + \xi_i - \xi_{i-1}.
\eeq
The parameter $\lambda$ controls the discretization protocol; $\lambda = 0$ corresponds to It\^o calculus, and $\lambda = 1/2$ is Stratonovich \cite{wio2013path}.
If we take $\lambda = 0$, it is clear that the diagonal entries of $\gamma^{-1}\dee \xi_i/\dee x_j$ are unit, and
$\calJ = \gamma$, but the time integrals in the action would need to be interpreted in the It\^o sense.
Setting $\lambda=1/2$, we can use the Stratonovich calculus with the usual interpretation of the integrals, at the cost of introducing a non-unit Jacobian. 
This gives
\eq 
\frac{1}{\gamma}\frac{\dee \xi_i}{\dee x_j} = \begin{cases}
1 +\frac{\Delta t}{2\gamma}\partial^2_x
V(x_i),\;\; j = i;\\
-1 +\frac{\Delta t}{2\gamma}\partial^2_x
V(x_{i-1}),\;\; j = i-1;\\
0\;\; {\rm otherwise}.
\end{cases}
\eeq 
This is an upper-triangular matrix, and its determinant is the product of its diagonal entries:
\eqa
\frac{1}{\gamma}\left|\frac{\dee \xi_i}{\dee x_j}\right| & = & \left( 1+\frac{\Delta t}{2\gamma}V''(x_0)\right)\left( 1+\frac{\Delta t}{2\gamma} V''(x_1)\right)\dots \left( 1+\frac{\Delta t}{2\gamma} V''(x_N)\right)+ \calO (\Delta t^2)\nonumber\\
& \to & \exp \;\sum_i \frac{\Delta t}{2\gamma} V''(x_i)\;\;{\rm as}\;\;\Delta t\to 0.
\eeqa
Thus, in the limit we obtain
\eq 
\calJ[x] = \gamma\exp\left( \frac1{2\gamma}\int_{t_0}^{t_1}V''(x(t))\dee t\right). 
\eeq 
Recall the path-independent term in Eq.~\eqref{eq:action}, the action:
\eq
\int_{t_0}^{t_1} 2\gamma V'\dot x\,\dee t  =  2\gamma \int_{x_0}^{x_1} V'\,\dee x = 2\gamma\left(V(x_1)-V(x_0)\right).\eeq
The above instead becomes
\eq
2\gamma \int_{x_0}^{x_1} V'\,\dee x = 2\gamma\left(V(x_1)-V(x_0)\right)+4\gamma D\int_{t_0}^{t_1} V''(x(t))\,\dee t
\eeq
if we interpret the integral in the It\^o sense.
This corresponds to $\calJ/\gamma = 1$, so the combined exponent is the same for both It\^o and Stratonovich interpretations, as it should be for an additive white noise process.
The procedure is analogous to evaluating an ordinary integral using Laplace's method:
\eq 
\int f(x)\exp\left(-\frac{g(x)}{M}\right)\dee x = \int\exp\left(-\frac1{M}\left( g(x) - M\log f(x)\right)\right)\dee x. 
\eeq 
The stationary points of $g$ differ only slightly from those of $g(x) - M\log f(x)$ as $M\to 0$, but in the functional case, this results in qualitatively different extremal paths. 


%

\end{document}